\documentclass[a4paper,fleqn,usenatbib]{mnras}

\usepackage{newtxtext,newtxmath}

\usepackage[T1]{fontenc}

\usepackage{graphicx}	
\usepackage{amsmath}	

\title[IR Spectra of V838 Mon in 2015 and 2022]{Infrared Spectroscopy of V838 Monocerotis in 2015 and 2022}

\author[T. R. Geballe et al.]{T. R. Geballe,$^{1}$\thanks{E-mail: tom.geballe@noirlab.edu(TRG)}
B. M. Kaminskiy$^{2}$,
D. P. K. Banerjee,$^{3}$
A. Evans,$^{4}$
Y. Pavlenko$^{2}$\thanks{Deceased.},
\newauthor
M. T. Rushton$^{5},$
M. Popescu,$^{5,6}$
S. P. S. Eyres$^{7}$
\\
$^{1}$Gemini Observatory/ NSF's NOIRLab, 670 N. A`ohoku Place, Hilo, HI 96720 USA\\
$^{2}$Main Astronomical Observatory, Academy of Sciences of the Ukraine, Golosiv Woods, UA-03680, Kyiv-127, Ukraine\\
$^{3}$Physical Research Laboratory, Ahmedabad, 380009, India\\
$^{4}$Astrophysics Group, Keele University, Keele, Staffordshire, ST5 5BG, United Kingdom\\
$^{5}$Astronomical Institute of the Romanian Academy, 5 Cu\c{t}itul de Argint, 040557 Bucharest, Romania\\
$^{6}$University of Craiova, Str. A. I. Cuza nr. 13, 200585 Craiova, Romania\\
$^{7}$Jeremiah Horrocks Institute, University of Central Lancashire, Preston PR1 2HE, United Kingdom
}
\date{Accepted XXX. Received YYY; in original form ZZZ}

\pubyear{2025}

\begin{document}
\label{firstpage}
\pagerange{\pageref{firstpage}--\pageref{lastpage}}

\maketitle

\begin{abstract}

We report medium-resolution 0.85$-$2.45~$\mu$m spectroscopy obtained in 2015 and 2022 and high resolution 2.27$-$2.39 $\mu$m and 4.59$-$4.77~$\mu$m  spectroscopy obtained in 2015 of V838 Monocerotis, along with modeling of the 0.85$-$2.45~$\mu$m spectrum.  V838 Mon underwent a series of eruptions and extreme brightenings in 2002, which are thought to have occured as a result of a stellar merger. The new spectra and modelling of them reveal a disturbed red giant photosphere that is probably continuing to contract and ejecta that are cooling and continuing to disperse at velocities up to 200 km~s$^{-1}$.

\end{abstract} 




\section{Introduction}

V838 Monocerotis \citep{bro02}, located at a distance of 6 kpc \citep{afs07,spa08,ort20} and coincident with a $V$=15 mag object unstudied prior to 2002, underwent a series of eruptions early in that year \citep{mun02a}. At one point it increased in visual brightness by almost ten magnitudes, briefly becoming optically more luminous than every known star in the Galaxy and creating a spectacular light echo observed at optical and far-infrared wavelengths \citep{bon03,ban06}. The eruptions produced unusual optical and infrared spectra \citep{mun02a,ban02}, the former of which contained broad P-Cygni profiles indicating ejecta traveling at several hundred km s$^{-1}$ \citep[e.g.,][]{kip04} and the latter of which also revealed ejection. On occasion V838 Mon produced highly unusual CO overtone band profiles near 2.3 $\mu$m  \citep{geb02,rus05a,geb07} and emission in overtone bands of SiO at 4 $\mu$m \citep{rus05b}. 

Following the 2002 eruptions V838 Mon quickly declined in visible brightness to close to its pre-eruption value \citep{cra03}. \citet{tyl11} found that by  2009 the optical spectrum of V838 Mon had lost virtually all signs of the violent events of 2002, and concluded that its spectrum approximated that of an M6 giant with an effective temperature of 3,270 K.  The decline at near-infrared wavelengths following the eruptions was much slower.
 
The most viable explanation for the striking photometric, spectroscopic, and kinematic evolution of V838 Mon and its ejecta appears to be a merger of a $\sim$8 ~M$_\odot$ star, whose spectral classification and evolutionary state prior to the merger are unclear, and a low mass star \citep{tyl05, tyl06, sok06}, in a process which occured in three stages over a period of two months \citep[see the light curves in][]{mun02a},  threw off  $\sim$0.1 M$_\odot$ of gas, and transformed the near-infrared spectrum into what was described by \citet{eva03} as that of an L supergiant. The L spectral type previously had only been used to describe main sequence stars and brown dwarfs.  High-resolution spectroscopy of the overtone and fundamental bands of CO in 2005-2006  \citep{geb07} revealed multiple ejected shells of gas spread out in velocity by over 200 km~s$^{-1}$ and a stellar photosphere that was contracting at 15 km~s$^{-1}$.

Mid-infrared interferometry  \citet{che14} performed approximately one decade following the violent events of 2002 resulted in the detection of thermal emission from dust over an extended region of several hundred a.u., consistent with the maximum expansion velocities that had been observed earlier in optical and infrared lines.  Their near-infrared interferometry revealed that the compact emission region observed with near-infrared interferometry in 2004 by \citet{lan05} had shrunk by 40\% in nearly a decade.  More recent optical speckle imaging and near-infrared interferometry by \citet{mob24} has revealed roughly bipolar asymmetries in the light distribution, which those authors tentatively interpret as jets. Given the small angular dimensions of these structures, extending $\sim$1 mas from the centre of V838 Mon,  which corresponds to a linear dimension of $\sim$1 $\times$ 10$^{14}$ cm on the plane of the sky, they appear to be unrelated to the emission by the high velocity material ejected over two decades ago. If they are jets, they may be revealing a different type of ejection than that caused by the 2002 merger events. As has been pointed out by all of these authors, the dimensions of the optical and near-infrared emissions are of the order of magnitude of the size of a late-type supergiant. The structures that \citet{mob24} detected could in part be asymmetrically distributed circumstellar material, or gas possibly associated with the continuing contraction of the disturbed photosphere of V838 Mon. 

The V838 Mon system is also known to contain a  B3V star \citep{wag02,cra03,mun02b}, which lies 0\farcs38 (225 AU) to the northwest of the merger \citep{kam21}. The presence of such a young star in the system suggests that the V838 Mon system is young and that elemental abundances in its component stars should not be far from solar. The B3V star has at times been partially or completely visually obscured from view, beginning 4.8 yr after the initial eruption, presumably by dust that formed in the ejecta \citep{mun07,tyl11}. We note that ejecta moving at 200 km s$^{-1}$  would take 5 years to travel 225 AU. The near equality of these time intervals suggests that the angular separation between V838 Mon and the B3V companion approximately corresponds to the physical separation. The B3V star also is believed to have excited the atomic emission lines observed in the optical spectrum of V838 Mon several years after the eruption \citep{bar06, mun07,tyl09} in ejected gas that passed close to the star.  

\citet{lii23} have recently reported optical - near-infrared photometry obtained since 2009, which shows that V838 Mon is becoming brighter and bluer.  They and  \citet{kam21} estimate that V838 Mon underwent an increase in photospheric temperature by 100$-$200~K between 2010 and 2020. They also suggest that these changes could be the result of a reduction in obscuration by circumstellar dust being formed in the ejecta from the merger. 

We and others monitored the infrared spectrum of V838 Mon at low and medium resolution at various times between 2002 and 2006. \citep{rus05a,lyn07,geb07}, and on occasion we obtained higher resolution spectra in selected wavelength intervals during that time \citep{rus05a,rus05b,geb07}. Since then, others have obtained spectra in the 1$-$2.5 $\mu$m interval, but only at medium spectral resolution (\citet{loe15} in 2008-2009; \citet{che14} in 2013-2014). Here we report additional medium-resolution near-infrared spectra obtained in 2015 and 2022 along with velocity-resolved spectra obtained in 2015 of portions of the CO fundamental and first overtone bands. When the high-resolution spectra were observed in 2015, it had been nearly a decade since velocity-resolved infrared spectra of the photosphere of V838 Mon and the ejecta were last obtained. 

 \begin{table*}
	\centering
	\caption{Observing Log.}
	\label{tab:log}
	\begin{tabular}{lcccccccc} 
		\hline
		UT Date & Tel/Instr & Wavelength  & Slit width& R & Exp. Time & Airmass & Telluric std & Airmass\\
		& & $\mu$m & arcsec & $\lambda/\Delta\lambda$ & sec & & & \\
		\hline
		20151130 & Gem N / GNIRS & 0.85-2.45 & 0\farcs30 & 1,400 & 64 s & 1.39 & HIP 34748 (A0V) & 1.27 \\
		20151210 & Gem N / GNIRS & 2.270-2.330 & 0\farcs10 & 18,800 & 360 s & 1.10 & HIP 30387 (A2IV) & 1.12 \\ 
		20151229 & Gem N / GNIRS & 2.325-2.385 & 0\farcs10 & 18,800 & 400 s & 1.10 & HIP 34748 (A0V) & 1.05 \\
		20151229 & Gem N / GNIRS & 4.59-4.77 & 0\farcs10 & 12,500 & 1,440 s & 1.14 & HIP 31583 (B8Ve) & 1.18 \\
		20220202 & IRTF/SpeX & 0.85-2.45 &  0\farcs30 & 2,000 & 56 s & 1.63 & HD53205 (B9IV-V) & 1.59 \\
		\hline
	\end{tabular}
\end{table*}
\section{Observations and Data Reduction}

Medium- and high-resolution infrared spectra of V838 Mon were obtained on three nights in late 2015 at the Frederick C. Gillett Gemini North Telescope, using the facility instrument GNIRS \citep{eli06}, for program GN-2015B-FT-20.  Pertinent information about the observations is presented in Table~\ref{tab:log}. The sky was clear on all three nights. Each spectrum was obtained using the standard stare / nod-along-slit observing mode.  The 0.85$-$2.45~$\mu$m (cross-dispersed) spectrum was reduced using the GNIRS reduction pipeline, which applies a flat field, removes bad pixels, wavelength-calibrates, and ratios by the spectrum of an A0V standard star, reduced in a similar fashion but with its \ion{H}{I} absorption lines removed using a Vega model spectrum. The single order high-resolution spectra covering 2.27$-$2.39~$\mu$m and 4.59$-$4.77~$\mu$m were reduced using a combination of IRAF \citep{tod86,tod93} and Figaro \citep{cur14} to perform the same steps, except for the $M$-band data, where in addition emission lines of \ion{H}{I} 7$-$5 at 4.6538~$\mu$m and 11$-$6 at 4.6726~$\mu$m) in the spectrum of the early type telluric standard star were removed manually prior to ratioing.  Wavelength-calibration utilized an argon lamp for the cross-dispersed spectrum and telluric absoption lines in the spectra of the standard stars for the single-order high resolution $K$- and $M$-band spectra. The uncertainties in these calibrations are $\sim$0.0002~$\mu$m for the medium-resolution cross-dispersed spectrum and $\sim$2 km s$^{-1}$ for the high-resolution spectra. 

An additional medium-resolution infrared 1$-$2.5~$\mu$m spectrum of V838 Mon was obtained in photometric conditions on UT 2022 February 2 at the NASA Infrared Telescope Facility for program 2022A04, and used the facility instrument SpeX \citep{ray03} in its cross-dispersed mode (see Table 1 for details).  Reduction utlilized the SpexTool pipeline \citep{cus04}. The spectrum of an B9 IV-V star, observed at similar elevations as V838 Mon, was employed for both removal of telluric lines and flux calibration. 

Continuum flux densities in the spectra should be viewed with some caution, as the diameters of seeing disks for V838 Mon and the telluric standards were larger than the slit widths. The random noise levels in all figures are small, because both V838 Mon and the standard stars are very bright; however it is possible that there are systematic effects in the ratioed spectra due to incomplete cancellation of strong telluric absorption features. In the medium-resolution spectra this is particularly true for the wavelength intervals 1.34$-$1.45 $\mu$m, 1.80$-$2.02 $\mu$m, and 2.40$-$2.45 $\mu$m  

\begin{figure}
\begin{center}
\includegraphics[width=8.5cm]{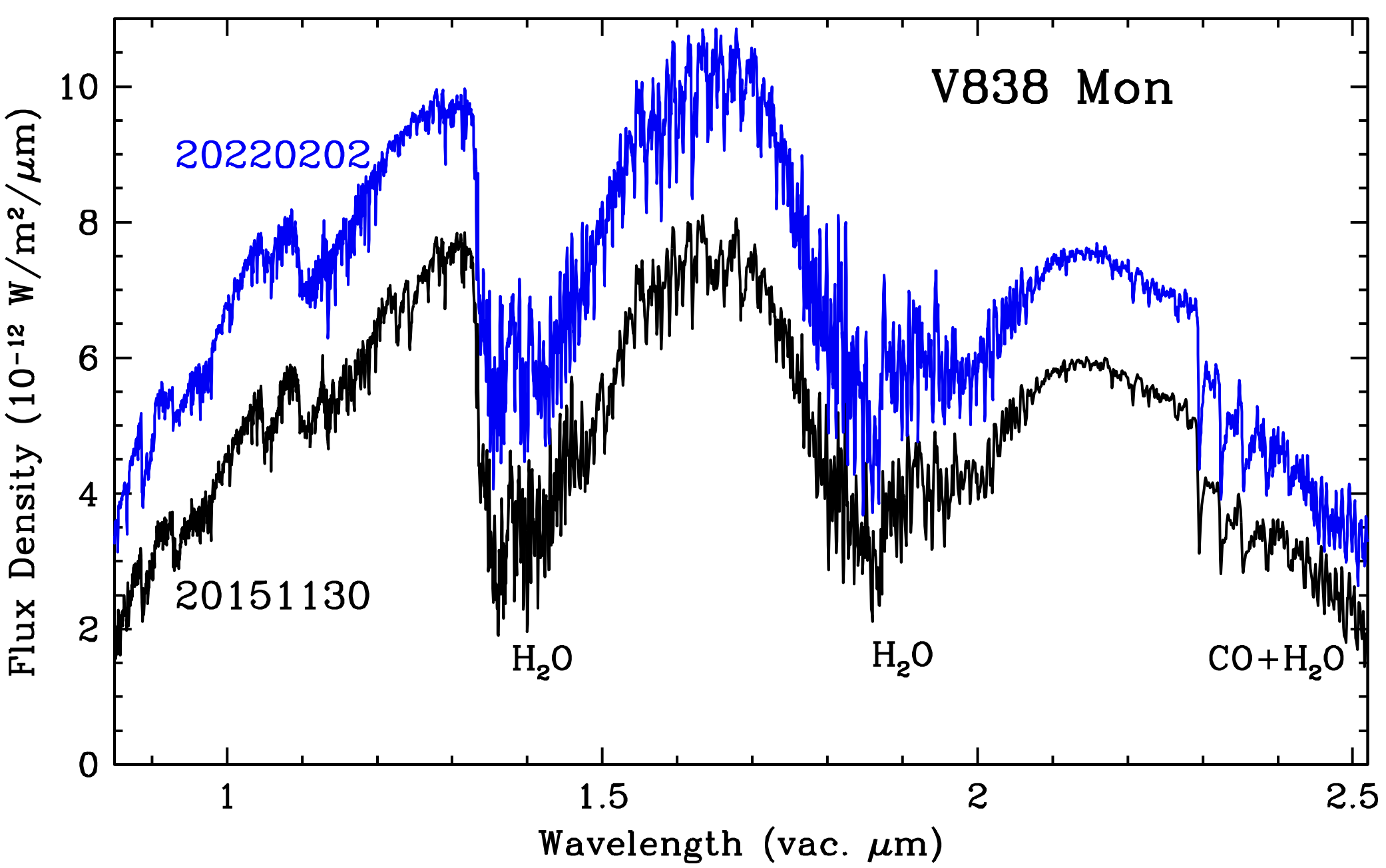}
\caption{Compressed view of 0.85$-$2.45~$\mu$m spectra of V838 Mon in 2015 (black) and 2022 (blue). Prominent broad molecular absorption bands as  are indicated.}
\label{fig:Fig2}
\end{center}
\end{figure}

\begin{figure*}
\begin{center}
\includegraphics[width=15cm]{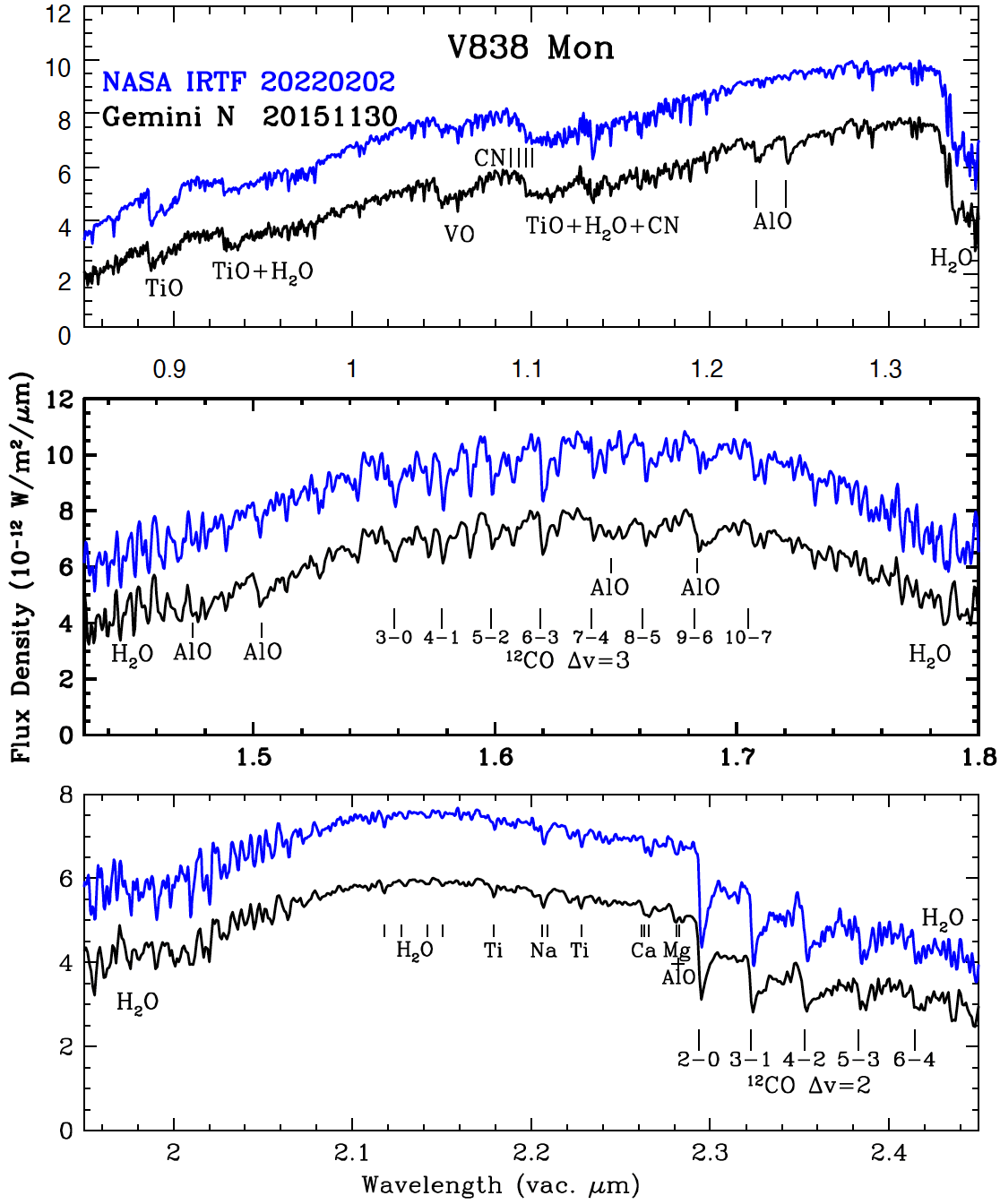}
\caption{More detailed view of the 0.85$-$2.45~$\mu$m spectra of V838 Mon in 2015 (black) and 2022 (blue). Prominent broad molecular absorption bands as well as numerous atomic and molecular lines are indicated.}
\label{fig:Fig2}
\end{center}
\end{figure*}

\begin{figure}
\begin{center}
\includegraphics[width=8.5cm]{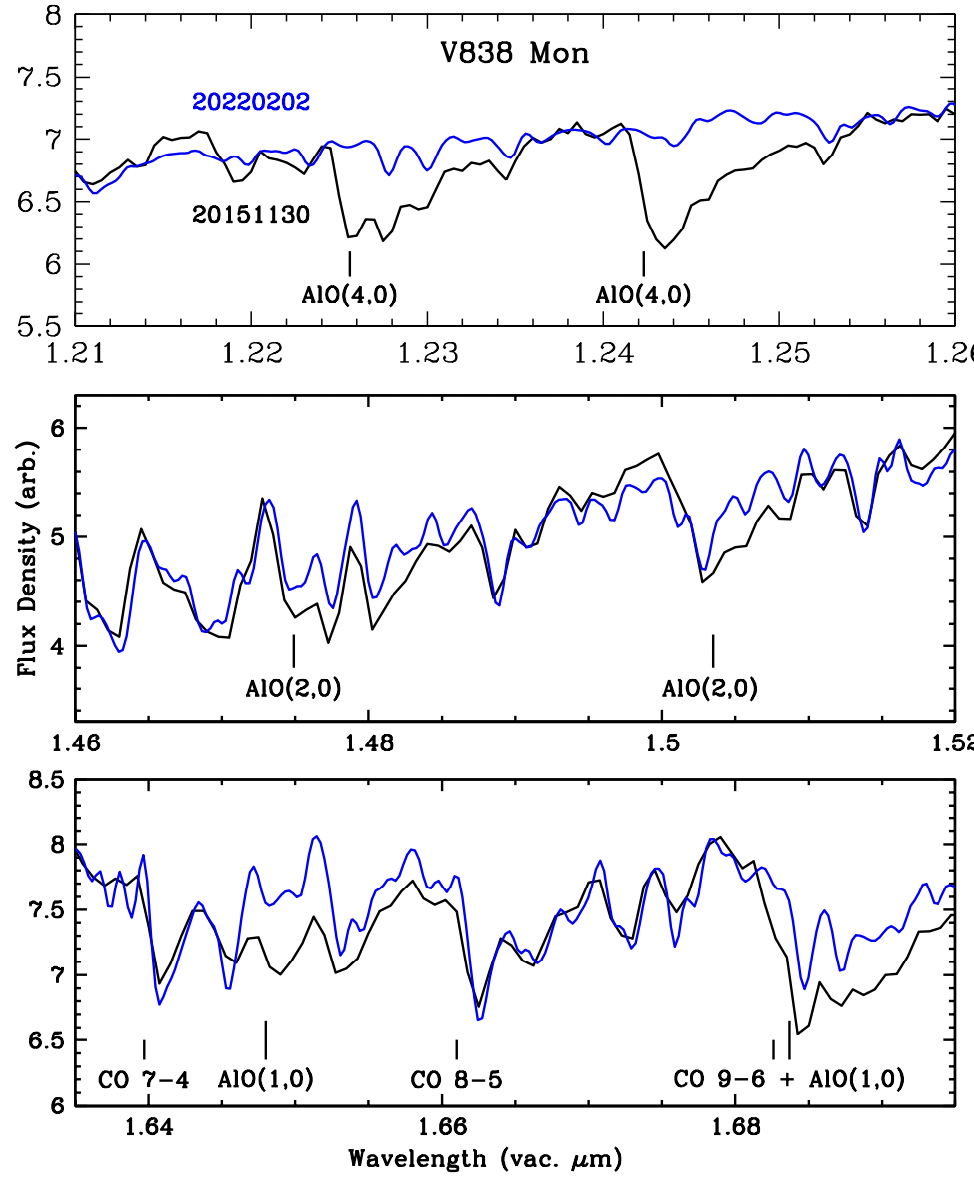}
\caption{Detailed view of medium-resolution spectra at wavelengths of AlO bands, The spectrum obtained in 2022 is scaled down slightly so that the continua in 2015 and 2022 approximately coincide. The heads of AlO absorption bands are indicated by the longer vertical lines; CO band heads in bottom panel  are indicated by shorter vertical lines. }
\label{fig:Fig3}
\end{center}
\end{figure}

\section{Results}

\subsection{Medium-resolution spectra}

Figures 1 and 2 show the observed medium-resolution 0.85$-$2.45~$\mu$m spectra obtained in 2015 and 2022. The spectra are remarkably similar in most respects, the most obvious difference being the increase in continuum brightness across the entire wavelength region, which despite the uncertainties in flux calibration we believe is real, as discussed below. The overall appearance of the spectrum from late 2015 (black traces), including both its absolute and relative peak flux densities in the $J$, $H$, and $K$ windows, is very similar to the spectrum observed by \citet{che14} two years earlier, in 2013 December.  Strong and broad bands of absorption by H$_2$O centred at $\sim$1.4~$\mu$m and $\sim$1.9~$\mu$m, as well as longward of $\sim$2.4~$\mu$m, dominate both spectra, with prominent first overtone absorption bands of CO also contributing deep absorption features longward of 2.3~$\mu$m.  The same dominant H$_{2}$O and CO absorptions are present in the 2022 spectrum.

Many additional absorption lines and bands can also be identified in both the 2015 and 2022 spectra; some of these are labeled in Figure~\ref{fig:Fig2}. At the shorter wavelengths (upper two panels) absorption bands of TiO, VO, CN, AlO, H$_2$O, and the second overtone bands of CO are present; together they cover much of the 0.9$-$1.7~$\mu$m interval and some of them are blended. Interestingly, the bands of AlO, which were first observed in V838 Mon in absorption in 2002 October \citep{eva03} and had been present in every reported spectrum of V838 Mon since then, are absent in the 2022 spectrum (see Figure~\ref{fig:Fig3} for more detailed views). The weakening of these bands during 2002-2005 had been noted by \citet{rus07}. The AlO bands have been found in only a handful of stars, all either red giants in advanced stages of evolution or eruptive variables; for additional information see \citet{ban12}.  Two other stars in which AlO band strengths have varied are known. One is the extremely cool asymptotic giant branch star IRAS 18530+0817. The $J$-band absorptions shown in the top panel of Figure~\ref{fig:Fig3} are present (also in absorption) in the spectrum obtained in 1995 by \citet{wal97}, but are absent in the 2008 spectrum of \citet{ban12}. The other is V4332 Sgr, in which the  AlO bands, which were observed in emission, greatly decreased in strength between 2005 and 2014 \citep{ban15}.

The bands of the CO second overtone (middle panel of Figure~\ref{fig:Fig2}) are strong up to $v$=8$-$5 and {may be} present up to $v$=10$-$7. However, the 9$-$6 and 10$-$7 band heads are blended with other features, adding uncertainty to this possibility.  If even higher band heads (longward of 1.7~$\mu$m) absorb, their distinctive spectral features are overwhelmed by absorption due mainly to H$_{2}$O, which extends from 1.7~$\mu$m to 2.1~$\mu$m.  At longer wavelengths (bottom panel of Figure~\ref{fig:Fig3}) only the aforementioned  CO and H$_2$O bands produce strong absorptions, but numerous weaker atomic and molecular lines can be discerned. 

The increase in continuum brightness between 2015 and 2022 is similar to what has been reported by \citet{lii23} based on infrared photometry in 2010 and 2020 \citep[photometry from the latter year obtained by][]{woo21}. Comparison of  continuum flux densities in our spectra at several wavelength intervals between strong absorption features indicates brightening ranging from 0.5 mag near 0.9 $\mu$m to 0.3 mag in the middle of the $K$ band.  \citet{lii23} suggested that decreasing reddening and/or a temperature increase of the cool supergiant could explain the infrared photometric changes they reported.  Our 2015 and 2022 spectra do not appear to provide evidence for a temperature increase of the cool supergiant's photosphere, and seem more consistent with a reduction in the column density of circumstellar dust on the sightline to the merged star. (Note that the slightly deeper narrow absorption features in the 2022 spectrum are the result of higher spectral resolving power for it than for the 2015 spectrum: 2,000 vs. 1400 in 2015).  Decreasing reddening due to lower column density of circumstellar dust could be the result of any of the following: radial expansion of the ejecta, reduction in the mass loss rate of the cool giant, and evaporation of recently created dust due to absorption of UV radiation from the nearby B3V companion. 

The reason for the disappearance of the AlO bands between 2015 and 2022 is unclear. The temperature of the photosphere of V838 Mon reported by \citet{tyl11} is probably too high for AlO to exist there; it has not been detected in normal supergiants with similar temperatures. The dissociation energy of AlO is $\sim$5 eV \citep{tyt67,hil73,dag75}, indicating that its abundance in cool stellar atmospheres could be highly sensitive to temperatures  of $\sim$2,000 K, but such temperatures are much lower than that of the photosphere of V838 Mon. The relative strengths of the continua in the $J$, $H$, and $K$ bands correspond to those of a 2,300 K blackbody, after dereddening the observed 0.85$-$2.45~$\mu$m spectrum for $E$($B-V$) = 0.85 (see below). In view of the much higher photospheric temperature, this suggests that the near-infrared spectrum of V838 Mon is a composite of the photospheric spectrum and the spectrum of much cooler gas and dust located well outside of the photosphere. It seems likely that the AlO bands arose in a circumstellar component and that the decreases in their strengths is related to changes in density and temperature of that component. Modeling of the 0.8$-$2.5~$\mu$m spectrum of V838 Mon is presented in Section 4.

\begin{figure}
\includegraphics[width=8.5cm]{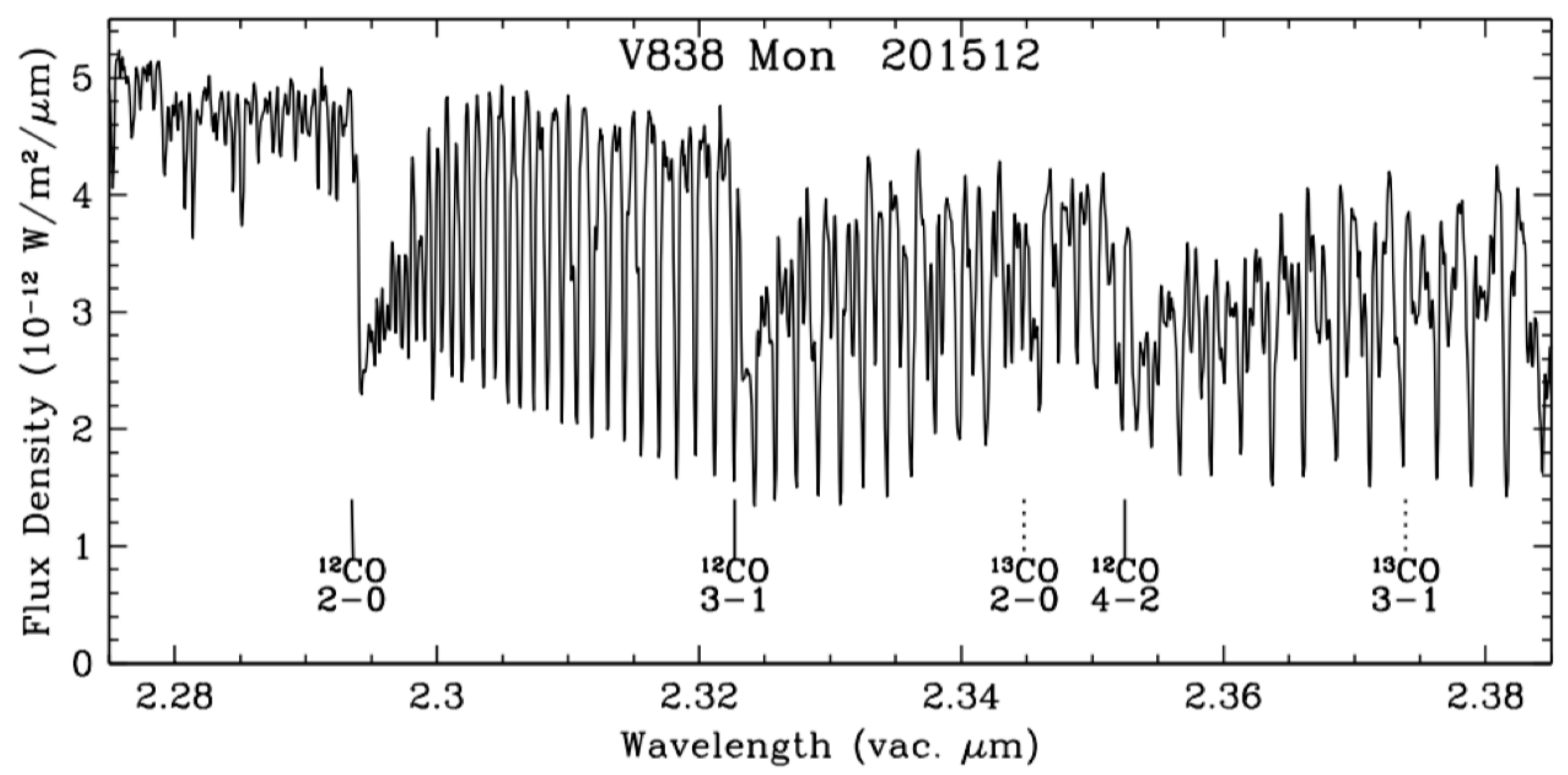}
\caption{High-resolution spectrum of V838 Mon in the region of the CO first overtone bands. The positions of the band heads of $^{12}$C$^{16}$O and $^{13}$C$^{16}$O are indicated.}
\label{fig:v838_co_k}
\end{figure}

\subsection{CO overtone band: high-resolution}

Figure~\ref{fig:v838_co_k} shows the high-resolution spectrum of a portion of the CO first overtone band.  Numerous individual absorption lines and the first three band heads of $^{12}$C$^{16}$O are clearly seen. The depths of the band heads are roughly comparable to those in red giants and supergiants \citep[e.g., see][]{wal96}.  All of the strong narrow absorptions in Figure~\ref{fig:v838_co_k} are individual 2$-$0 transitions of $^{12}$C$^{16}$O. Because their depths are comparable to the depth of the 2$-$0 band head, where many lines overlap, they must be optically thick. 

{At  wavelengths shorter than $\sim$2.33 $\mu$m, where 2$-$0 band lines corresponding to transitions between high rotational levels occur, the 2015 spectrum closely resembles the high-resolution spectrum obtained in 2006 by \citet[][see their Fig. 3]{geb07}. However, at longer wavelegths  (2.33$-$2.35 $\mu$m) the appearances of the spectra are noticeably different.  This is due to the large changes between these two epochs in the profiles of lines corresponding to vibration-rotation transitions in lower rotational levels, as discussed in the following section.}

There is no obvious evidence in this spectrum for absorption due to  $^{13}$C$^{16}$O, indicating that the isotopic ratio $^{12}$C/$^{13}$C is much higher than observed in evolved oxygen-rich red giants \citep[e.g., see][]{mil09}. This is not surprising, in that the event thought to have transformed V838 Mon into a cool supergiant is unrelated to the internal processes that change the surface and circumstellar isotopic abundance ratios in evolved stars. From their earlier high resolution spectra of CO, \citet{geb07} tentatively estimated $^{12}$C/$^{13}$C $\sim$ 100.

\begin{figure}
\begin{center}
\includegraphics[width=7.0cm]{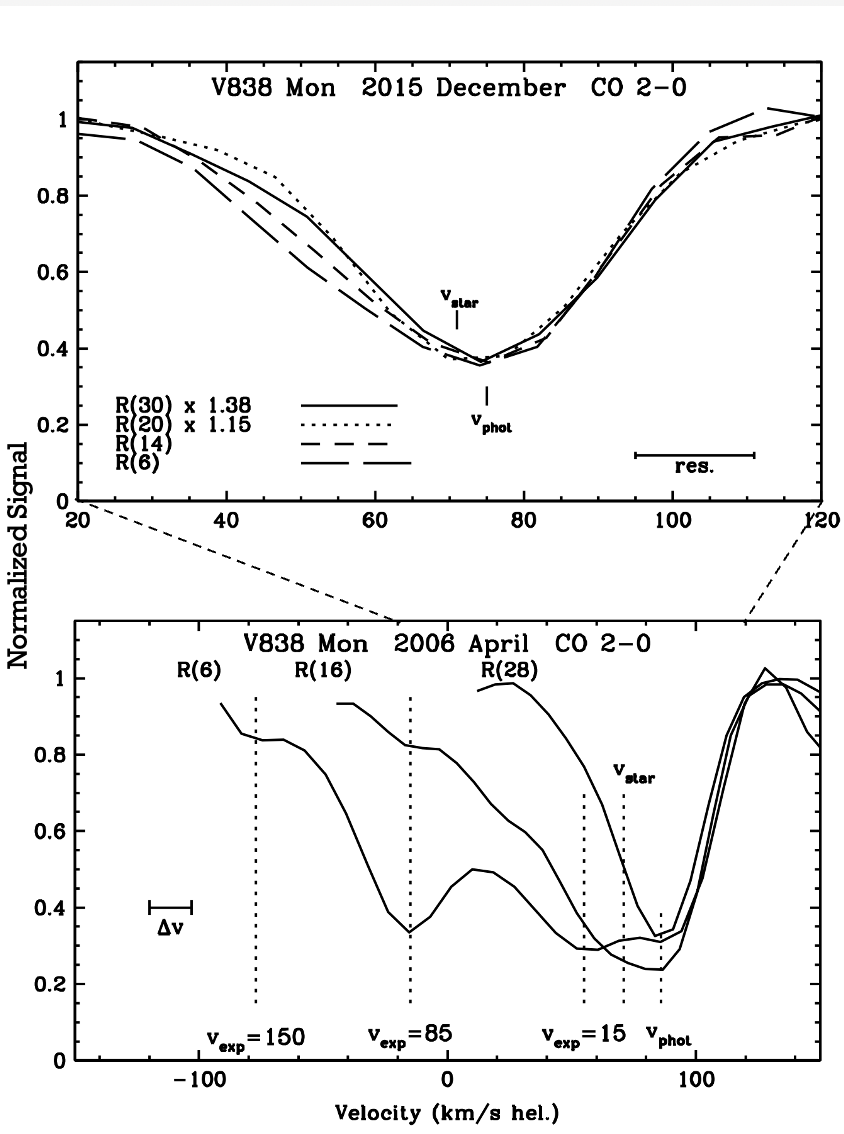}
\caption{{Velocity profiles of selected lines of the 2$-$0 band of $^{12}$C$^{16}$O in 2015 December (upper plot) and in 2006 April (lower plot, using the data in Figure 4 of \citet{geb07}). The velocity range shown for the 2006 profiles is three times that of the 2015 plot, as indicated. In the 2015 plot the depths of the profiles have been scaled by the factors shown and the velocities shifted by up to $\pm$2~km~s$^{-1}$ to align the velocities of peak absorption and the more positive velocity sides of the profiles. The stellar radial velocity of 71 km s$^{-1}$ \citep{tyl11} and estimated velocities of the stellar photosphere in 2005 and 2015 are indicated, as are expansion velocities of peak absorption in the ejected gas in the 2005 data. The velocity resolutions (16 km s$^{-1}$ in 2015 and 17 km s$^{-1}$ in 2005) are indicated.}}
\label{fig:v838_k_co_profiles}
\end{center}
\end{figure}

\begin{figure*}
\includegraphics[width=15.0cm]{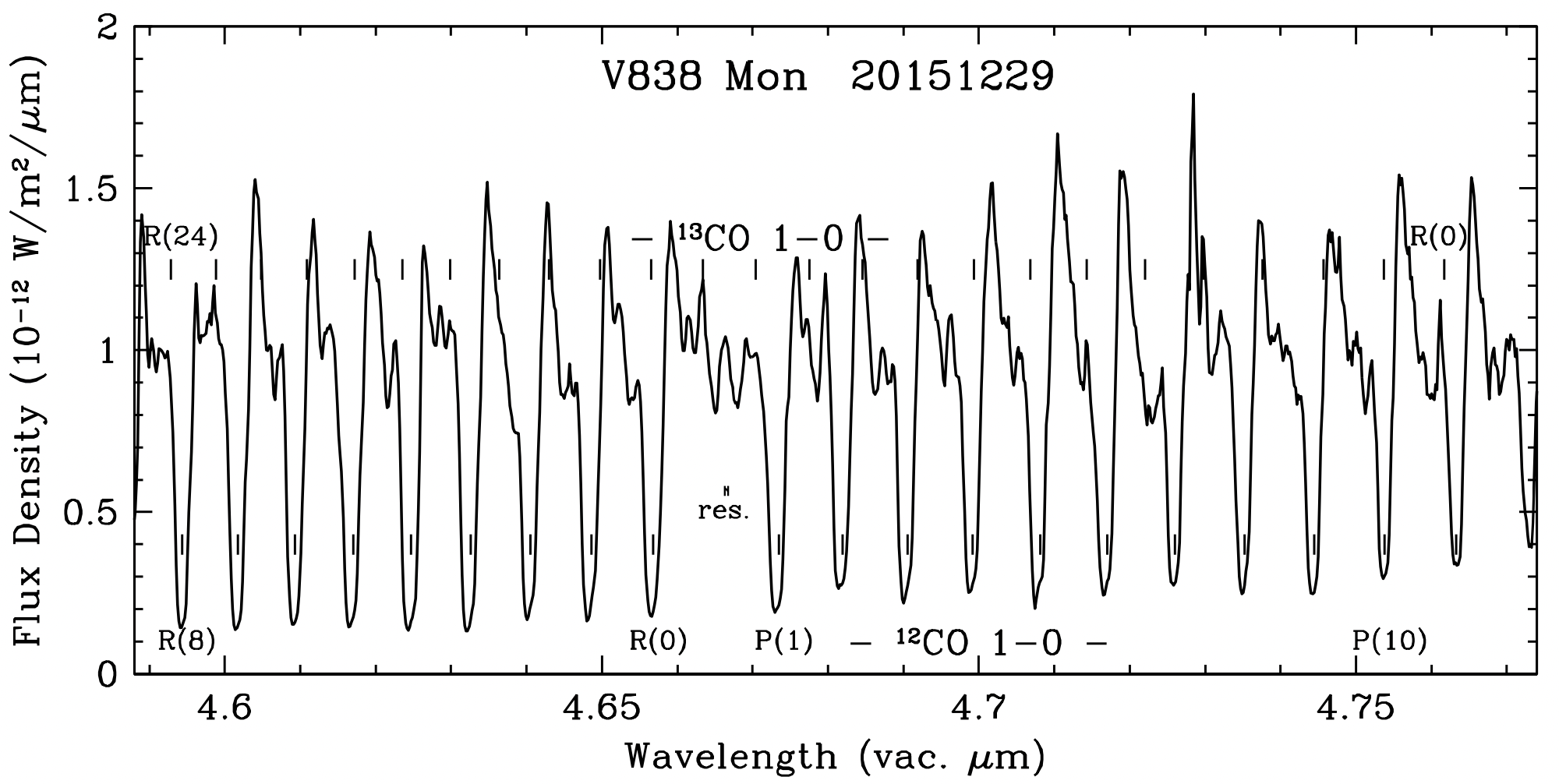}
\caption{High-resolution spectrum of V 383 Mon of a portion of the CO fundamental band.  Rest wavelengths of the lines are indicated by vertical tick marks, at bottom for $^{12}$C$^{16}$O and at top for $^{13}$C$^{16}$O (which was not detected). Some of the transitions are labelled.}
\label{fig:v838_co_m}
\end{figure*}

{Figure~\ref{fig:v838_k_co_profiles} provides detailed views of representative CO overtone lines from 2015, as well as previously reported profiles of lines observed in 2006. The upper plot} shows the velocity profiles in 2015 of four 2$-$0 band lines, $R$(6), $R$(14), $R$(20), and $R$(30), which orginate from rotational levels of the ground vibrational state having a wide range of energies (116, 580, 1150, and 2,555~K, respectively). These lines were chosen for examination because their profiles are uncontaminated by those of other CO lines. It is possible, however, that their profiles are affected by weak lines of other species. The velocity of peak absorption in each of these 2$-$0 lines is 75~$\pm$~3 km~s$^{-1}$(heliocentric), which is redshifted by 4 $\pm$~3 km s$^{-1}$ from the heliocentric velocity of V838 Mon deduced from optical spectroscopy \citep{tyl11} and from observations of SiO masers \citep{deg05,cla07}. The observed full widths at half maximum (FWHMs) of the two narrowest lines, $R$(20) and $R$(30), are 38 km~s$^{-1}$, indicating intrinsic FWHMs of $\sim$30 km s$^{-1}$  The lower state energies of these two lines are considerably higher than those of the $R$(6) and $R(14)$ lines and are at sufficiently high energy that they must be formed within, or close to, the photosphere of the supergiant. 

In principle, the observed FWHMs of the $R$(20) and $R$(30) lines could be due to rotation and/or to turbulence.  Little is known about rotational velocities in cool supergiants, but there is a general expectation that they are very low due to the large radii of these stars, and therefore that the observed linewidths in them are due largely to turbulence \citep{kaf79}. Turbulent FWHMs of $\sim$30 km s$^{-1}$ are not unusual for red supergiants \citep{jos07}. As can be seen in Figure~\ref{fig:v838_k_co_profiles}, the $R$(6) and $R$(14) lines are slightly more asymmetric  than the two higher excitation lines, with broader blueshifted absorption. The logical interpretation is that the extra absorption arises in cooler gas outside of the photosphere that is still expanding radially, at 20$-$40 km s$^{-1}$.

{The CO overtone line profiles from 2015  December are profoundly different from those observed nearly a decade earlier \citep{geb07}, which are shown in the lower plot of Figure~\ref{fig:v838_k_co_profiles}}. {First, while none of the profiles observed in 2015 possesses extended blueshifted wings,} in 2006 the two $v$=2$-$0 absorption lines from rotational levels close to or identical to those of the $R$(6) and $R$(14) transitions have extremely broad blueshifted absorption, with identifiable components centred at $-$15, $-$85, and $-$150 km s$^{-1}$ relative to the stellar radial velocity. {Second, although} the shape of the $R$(28) absorption line from 2006 is similar to {that of the $R$(30) line in 2015}, its absorption peak was redshifted by +15 km s$^{-1}$ from the stellar velocity, whereas the {mean displacement of the 2015 absorption peaks} from the stellar velocity is only +4 $\pm$~3 km s$^{-1}$. \citet{geb07} concluded  from the {2006} spectra that the outer layers of the photosphere where the CO lines are formed were contracting. The current much smaller redshift together with the blue asymmetry in even the highest excitation lines in Figure~\ref{fig:v838_k_co_profiles} imply that the contraction has slowed considerably, but that settling of the outer layers of the stellar {atmosphere} is probably not completed. The absence of highly blueshifted components of the 2$-$0 band lines in 2015 that were detected in 2006 indicates that the line of sight column densities in the high velocity gas are much lower than a decade ago, as would be qualitatively expected due to rapid expansion of the ejecta that produced those components. 

\subsection{CO fundamental band}

The 4.59$-$4.77-$\mu$m spectrum of V838 Mon (Fig.~\ref{fig:v838_co_m}) obtained in 2015 is dominated by very broad lines of the fundamental ($v$=1$-$0) vibration-rotation band of $^{12}$C$^{16}$O, as it was when last observed in 2005 \citep[][see their Fig. 5]{geb07}. All of the CO lines in the 2015 spectrum have P Cygni profiles and the full range of absorption and emission for each line is nearly as broad as the separations of adjacent  lines.  P Cygni profiles may also have been present in the 2005 spectrum of the fundamental band, but the presence of line emission at that time is less obvious as the spectrum was much more badly contaminated by lines of other species.

Despite the continued expansion of the ejecta, the absorption components of the $^{12}$C$^{16}$O fundamental band lines in 2015 are heavily saturated, as they were a decade ago.  The continuum being absorbed by the CO must be predominantly thermal emission from warm circumstellar dust rather than continuum from the stellar photosphere. This is because (1) the dereddened $JHK$ magnitudes of V838 Mon in 2015 fit those of a $\sim$2,300~K blackbody, and (2) the temperature of the stellar photosphere is in excess of 3,000 K \citep{tyl11}. In either case, assuming that the observed $K$-band continuum comes entirely from the star, the $M$-band continuum from the star would contribute at most only one-fourth of the observed 4.7 $\mu$m continuum. In addition, any stellar continuum contributing at $\sim$4.7~$\mu$m would be depressed by photospheric absorption from the CO fundamental lines, as well as by absorption from circumstellar dust lying outside of the photosphere but inside the gas that produces the 4.7 $\mu$m CO lines.

As can be seen in Figure~\ref{fig:v838_co_m},  weak absorption lines of other species are present in between the strong $^{12}$C$^{16}$O absorption lines. Some of these weak lines may coincide with some of the CO P Cygni emissions and thus be responsible for the observed line-to-line variations the P Cygni emission strengths. In view of the strong absorption by H$_{2}$O in the near-infrared spectrum, we suspect that most if not all of the additional absorption lines are due to H$_{2}$O, but we have not attempted to identify individual transitions. Although lines of the 1$-$0 band of $^{13}$C$^{16}$O correspond to some of these weaker features, others do not and thus, similar to the overtone band, we are unable to identify lines of this isotopic species. We conclude that the lower limit on $^{12}$C/$^{13}$C from the fundamental band also is consistent with a solar-type value for that ratio.

\begin{figure}
\includegraphics[width=7.0cm]{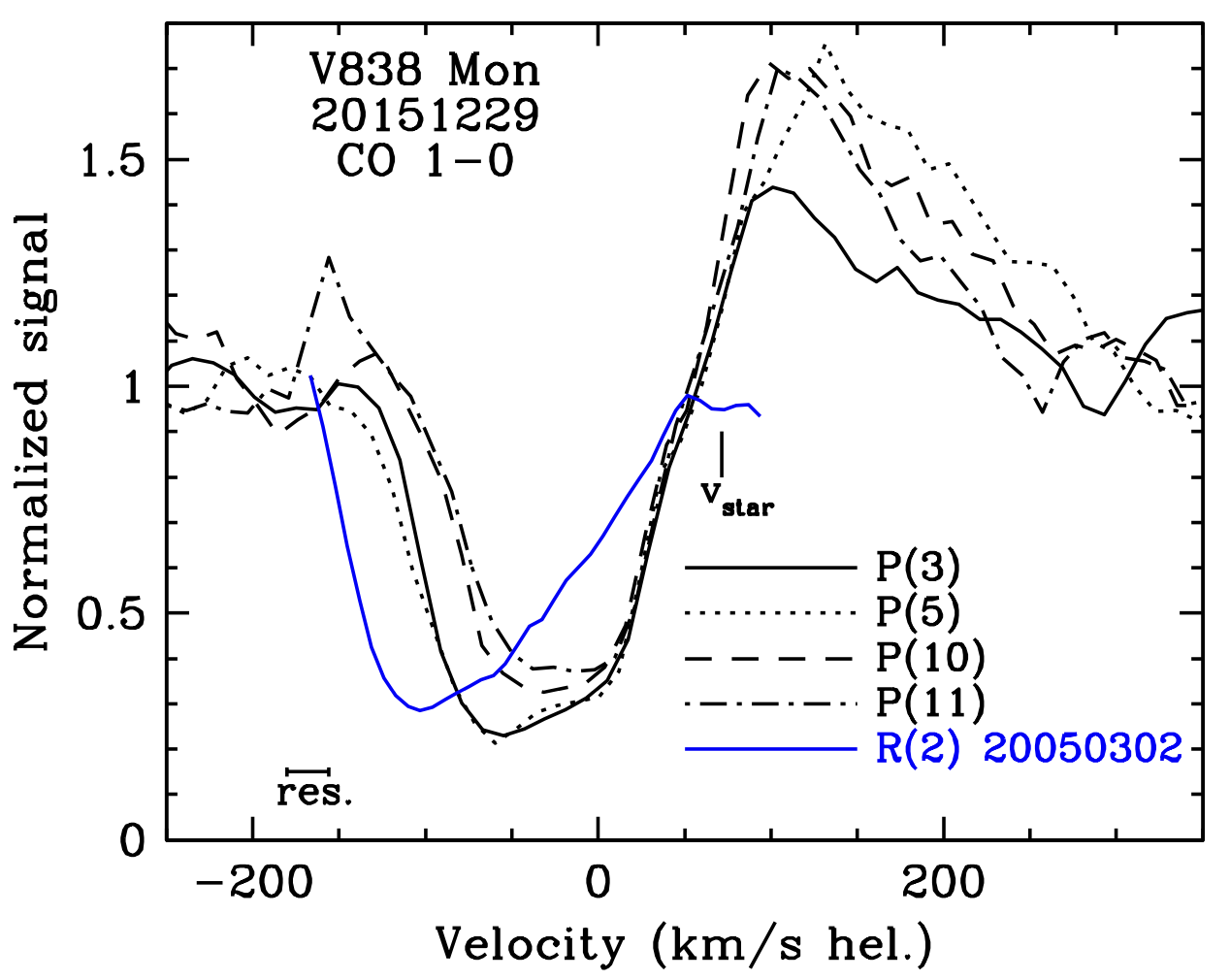}
\caption{Velocity profiles of four $P$-branch lines of the 1$-$0 band of $^{12}$C$^{16}$O observed in 2015 (black curves) and one line from the 1$-$0~$R$ branch from \citet{geb07} in 2005 (blue curve).  Small scaling adjustments ($\sim$5\%) have been applied to the $P$-branch profiles in Fig.~\ref{fig:v838_co_m} to align the continuum levels. The stellar radial velocity is denoted by a short vertical line. The velocity resolution (24 km s$^{-1}$) is shown at bottom left.}
\label{fig:v838_co_m_profiles}
\end{figure}

\begin{figure*}
\begin{center}
\includegraphics[width=13.0cm]{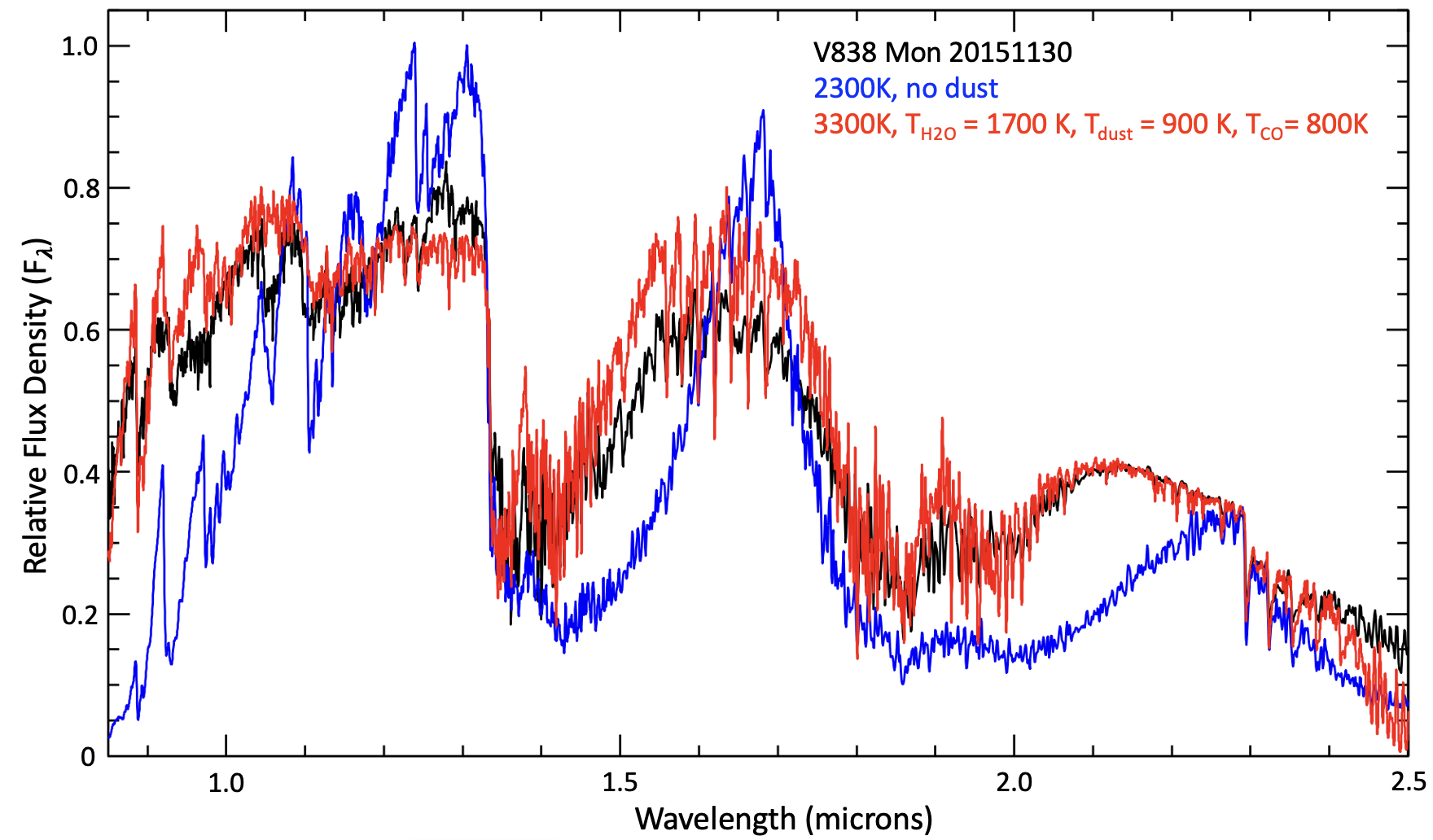}
\caption{Observed dereddened 0.8$-$2.5~$\mu$m spectrum of V838 Mon. (black trace) and two model spectra with log g = 0 and solar metallicities. 
The model shown in blue is for a 2,300 K supergiant; the model shown in red, is for a 3,300 K supergiant photosphere with emission from dust at 900 K, scaled to veil the CO first overtone bands by the observed amount, and 1,700 K circumstellar gas containing H$_{2}$O to match the strengths of the 1.4 and 1.9 $\mu$m bands.}
\label{fig:Fig8}
\end{center}
\end{figure*}

The P Cygni velocity profiles of four $P$-branch lines of CO are plotted in Figure~\ref{fig:v838_co_m_profiles}. The line absorption and the P Cygni emission each extend (in opposite directions) approximately 200 km s$^{-1}$ from the stellar radial velocity, with lines from lower energy levels having somewhat more highly blueshifted absorption. This is qualitatively as expected, for several reasons. The more distant absorbing CO should have both a higher expansion velocity and be cooler than CO nearer to the star. In addition, the more distant and higher velocity CO would be expected to be in lower density gas that would be less able to maintain LTE populations in the higher $J$ levels of the ground vibational states \citep{yan10}.

Figure~\ref{fig:v838_co_m_profiles} also includes the velocity profile of the 1$-$0~$R$(2) line observed ten years earlier, as reported by \cite{geb07}. Although the velocity ranges of the line absorption are similar in 2005 and 2015, the velocity of maximum line absorption, which corresponded to an expansion velocity of $\sim$160 km s$^{-1}$ in 2005, had decreased to $\sim$130 km s$^{-1}$ in 2015.  

The wide absorption troughs in the CO profiles observed in 2015 imply that expansion of the ejecta was continuing over a wide range of velocities and therefore over a wide range of distances from the merged star.  Near the blueshifted edges of the absorptions both the higher and lower $J$ lines in  Figure~\ref{fig:v838_co_m_profiles} are unsaturated and their relative strengths allow one to estimate the  rotational temperature (T$_{\rm rot}$) of the CO (which should be the same as the gas kinetic temperature). At $v_{\rm hel}$ =  $-100$ km s$^{-1}$  (corresponding to an expansion velocity of 170 km s$^{-1}$ we obtain T$_{\rm rot}$ $\sim$ 100 K. At lower expansion velocities, the absorption depths are comparable but the lines are saturated; assuming that their optical depths are identical the gas temperature is $\sim$500 K.            

Because the temperature-equivalent energies of the $v$=1 levels producing the redshifted P Cygni emissions in Figure~\ref{fig:v838_co_m_profiles} are $\geq$3,000 K, the emissions are unlikely to be the result of thermal excitation of rotational levels in the $v$=1 state, as the temperature-equivalent energies of the those $v$=1 levels exceed 3,000 K and their lifetimes against spontaneous emission are short. We conclude that the P Cygni emission is the result of resonant scattering of 4.7$\mu$m radiation mainly from dust mixed with and interior to the emitting CO, surrounding the merged star. 

\section{Comparison with Model Spectra}

\subsection{Medium resolution}

Given the unsettled state of V838 Mon in 2015 and 2022, clearly still recovering from the violent events of early 2002, no standard atmospheric model is likely to accurately reproduce the near-infrared spectrum, even though, when viewed at moderate spectral resolution, the overall shape of the near-infrared spectrum has remained roughly constant during the last $\sim$15 years \citep[Fig. 2, see also][]{che14,loe15}. The $M$-band spectrum in Figure 6 demonstrates that in the thermal infrared the photosphere of V838 Mon is completely obscured (indicating that at shorter infrared wavelengths veiling of the spectrum may be significant). However, many of the spectral features observed in the 1$-$2.5~$\mu$m interval are the same ones seen in the photospheres of cool late-type giants and supergiants, e.g. the CO first and second overtone bands, several metal oxides, CN, and possibly H$_{2}$O as well (depending on the photospheric temperature). Thus it is initially instructive to compare the 1$-$2.5~$\mu$m spectrum of V838 Mon to that of a model stellar photosphere of similar luminosity and temperature. 

The relative continuum flux densities of V838 Mon in the $J$, $H$, and $K$ bands (i.e., near 1.25, 1.65, and 2.2 $\mu$m) can be approximated by those of a 2,300 K blackbody. At near-infrared wavelengths the existence and spectral dependence of veiling by emission from circumstellar dust is unknown. Ignoring that possibility initially, we generated a model spectrum of a 2,300 K supergiant with solar abundances for comparison with the 2015 dereddened spectrum of V838 Mon. The $K$ magnitude of V838 Mon at that time was $\sim$4.4. For comparison, at a distance of 6 kpc, the $K$ magnitude of $\alpha$ Orionis (M2 I, T $\approx$ 3,600 K, d = 580 pc, $K$=-4.05) would be 1.0, twenty times brighter than V838 Mon, and Mira at minimum visible brightness / lowest temperature  (T~$\sim$~2,000 K, d = 128 pc, $K~\sim$~2.2) would have $K$ = 6.2, one-fifth as bright as V38 Mon. These comparisons suggest that coarsely characterizing V838 Mon in 2015 as a 2,300 K supergiant is a reasonable first step.

{The synthetic spectra were computed using the WITA6 code \citep{pav97} within a classical framework: LTE, hydrostatic equilibrium, and a one-dimensional model atmosphere without sources and sinks of energy. All abundances used are solar. 
Models of the atmosphere were computed using the SAM12 code \citep{pav03}, which is a modification of ATLAS12.
To model the spectral lines formed in the shell, we used the approach described in Section 8.2 of  \citet{eva19}. Atomic lines were taken from VALD3 database \citep{rya15}. Data for the main molecular species, CO, H$_{2}$O, TiO, and AlO, are respectively from \citet{goo94}, \citet{barb06}, \citet{ple98}, and \citet{pat15}.}

The match of this model spectrum to the observed spectrum in 2015, shown in Figure 8, is very poor.  (It would also be poorly matched to the 2022 spectrum.) One of the most glaring discrepancies is the difference in strengths of the absorption bands of H$_{2}$O at 1.35$-$1.5  
$\mu$m,  1.8$-$2.2 $\mu$m, and longward of 2.4 $\mu$m; in the model they are far too strong. This suggests that the assumed photospheric temperature of  2,300 K needs to be substantially increased in order to reduce the H$_{2}$O abundance. There are other discrepancies as well: near 1.0 $\mu$m the model spectrum is far too weak and the molecular bands far too strong. The overtone CO bands near 2.3 $\mu$m are also too strong.  The first of these also suggests that the photospheric temperature should be increased. The second of these discrepacies could be interpreted as implying a much lower carbon abundance, but this is highly unlikely, given the youth of the V838 Mon system. However, another factor could be involved. The observed $M$-band spectra, which shows a dust continuum and no indication of photospheric CO absorption, such as is seen in red supergiants, suggests that veiling by dust may be significant at shorter wavelengths, and in particular so in the $K$ band. 

\begin{figure*}
\begin{center}
\includegraphics[width=15cm]{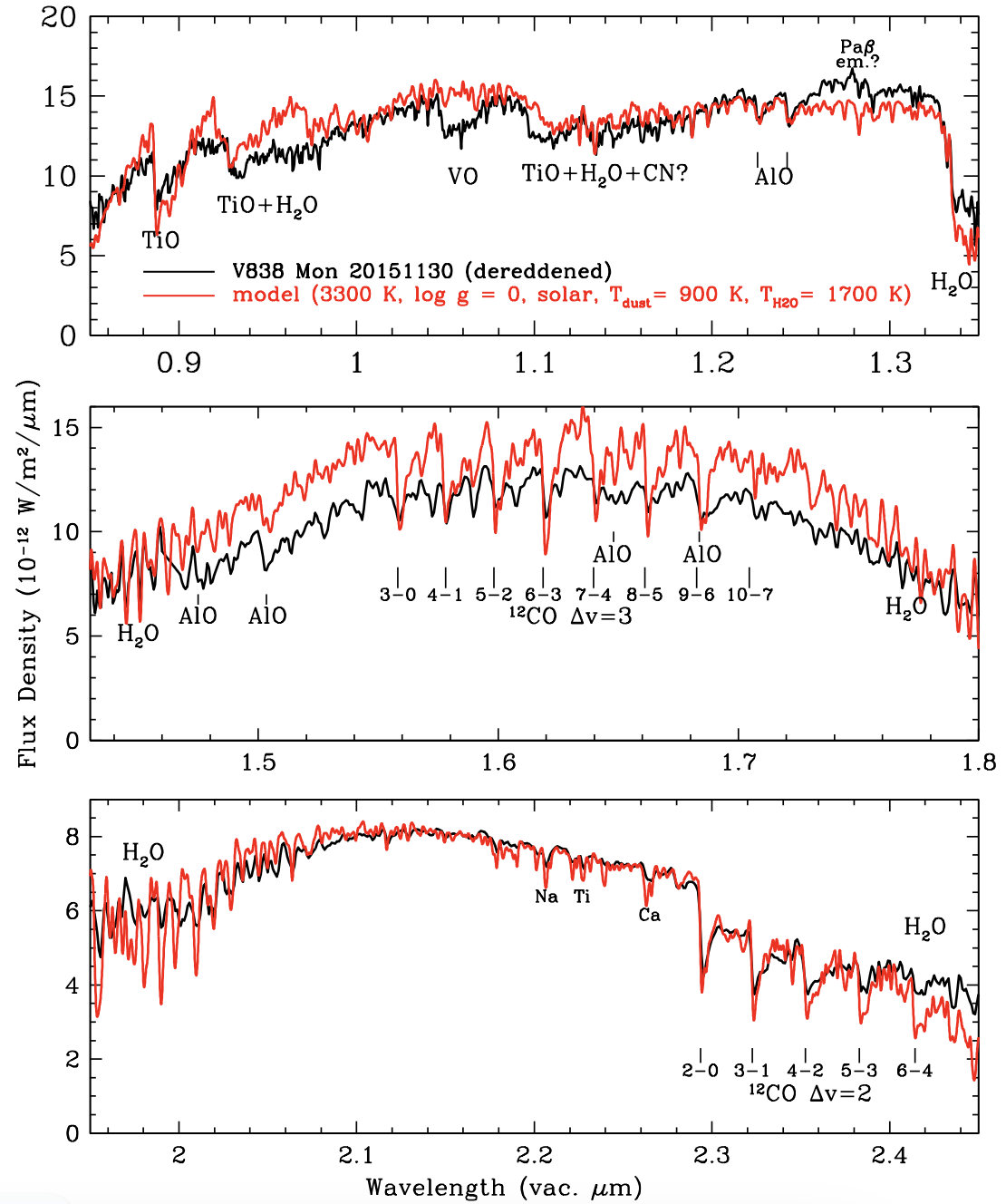}
\caption{Detailed view  of the 0.85$-$2.45~$\mu$m model spectrum (red) in Fig. 8.and the 2015 spectrum of  V838 Mon (black). Prominent broad molecular absorption bands as well as numerous atomic and molecular lines are indicated.}
\label{fig:Fig9}
\end{center}
\end{figure*}

Therefore, we created hybrid models incorporating more than one spectrum-producing component. One such model, also shown  in Figure 8, includes four components: a 3,300 K photospheric spectrum (which matches the temperature found by \citet{tyl11} from the optical spectrum and nearly eliminates absorption by photospheric H$_{2}$O); gray-body emission from circumstellar shell of dust at T = 900 K to veil the long wavelength portion of the K window and weaken the CO overtone absorption bands; a circumstellar H$_{2}$O at 1,700 K (inside the dust shell) to match the observed strengths of its 1$-$2.5~$\mu$m absorption bands; and circumstellar CO at T = 800 K to account for the fundamental band absorption lines.

This hybrid model is a considerably better fit to the observed spectrum than a pure photospheric model. However, it is  not the last word in fitting the complex spectrum of V838 Mon.  In addition to mismatched absorption band strengths (e.g., at 1.15 $\mu$m  and in the strong H$_{2}$O bands at 1.45 $\mu$m and 2.0 $\mu$m) there are significant differences in the continuum level in several spectral intervals, e.g. 1.50$-$1.75~$\mu$m.  The latter could be due, for example, to wavelength-dependent dust emissivity. In general, however, the model is remarkably close to the observed spectrum. This indicates that although its spectrum may have been described as that of an ``L supergiant" photosphere shortly after the 2002 eruption \citep{eva03}, multiple components, including a photosphere with a much higher temperature, similar to that found by \citet{tyl11}, one or more components that are much cooler and produce continuum emission from dust, and one or more components that produce strong absorption by circumstellar H$_{2}$O, CO, and possibly other molecular species, are required to roughly approximate more recent optical-infrared spectra of V838 Mon. In reality, one expects the various circumstellar components to be present over wide ranges of temperatures and densities. 

Figure 9 is a more detailed comparison between the hybrid model and the observed spectrum.  To roughly account for the diffference in CO linewidths in the photospere and in the ejected gas, the model incorporated extra broadening for the lines formed in the shell relative to those of the photospheric CO. 

AlO line absorption from circumstellar gas has been included in the model as a separate  and independent component. From their analysis of spectra of  optical AlO B-X bands near 5000 {\AA} observed in 2005 and 2009, \citet{kam09} and \citet{tyl11} derived an excitation temperature of $\sim$1500 K at both times. Our modeling of the AlO A-X features near 1.23 $\mu$m, observed in 2015, when absorption was only present near the band heads, results in a  much lower temperature, $\sim$ 200 K (with even a 300 K excitation temperature clearly ruled out). Thus, the gas producing the AlO bands  was much cooler in 2015 than it was at the earlier times. As mentioned previously, the infrared AlO bands are completely absent in our 2022 spectrum, implying further cooling occured after 2015. 

It was not possible to model the AlO lines near 1.5 and 1.66 $\mu$m due to the overall  poor fit to the spectrum in those wavelength regions as well as contamation by other strong absorbers (i.e., absorption by the CO second overtone near 1.66 $\mu$m). It does appear that the AlO bands near 1.5 $\mu$m were present in 2015, and like the 1.23 $\mu$m bands, were absent in 2022.  \citet{tyl11} discuss AlO formation and also point out that in cool stellar atmospheres Al is mostly atomic. Hence it is not suprising that the distribution of AlO in the photosphere and circumstellar gas differs from those of more tightly bound molecules such as CO.

There are numerous examples of narrow lines that are noticeably stronger in the model than in the observed spectra. Detailed examination of these is beyond the scope of this paper. {In Figure 9 three examples of this are indicated: atomic lines of Na, Ti, and Ca in the 2.2$-$2.3~$\mu$m range.} Possible explanations include deviations from solar elemental abundances, perhaps combined with additional veiling. 

\subsection{High resolution: $^{12}$C/$^{13}$C}

Modelling of the high-resolution CO overtone spectrum in Figure 4 with the hybrid model parameters failed to reveal evidence for a contribution by lines of  $^{13}$C$^{16}$O. Comparison of the model CO spectrum with the observed spectrum is severely hampered by interference in the observed spectrum by lines of H$_{2}$O, especially those lines located near the band heads of $^{13}$C$^{16}$O, which in normal red giants are the most easily recognized $K$-band spectral features of that CO isotopomer.  The broad absorption troughs of the 2$-$0 lines of $^{12}$C$^{16}$O from the lowest rotational levels (note the profile of the 1$-$0 $R$(2) line in Figure 7), which occur close to the 2$-$0 band head of $^{13}$C$^{16}$O near 2.325 $\mu$m are an additional complication.  We can only conclude that it is highly unlikely that $^{12}$C/$^{13}$C is significantly lower than $\sim$100; the model fits actually suggest that it is considerably higher than that value. 

\section {Conclusions}

We have presented medium-resolution near-infrared spectra of V838 Monocerotis in 2015 and 2022, roughly 13 and 20 years, respectively, after its cataclysmic outbursts in 2002. We also have presented high-resolution spectra of  V838 Mon in a portion of the CO overtone band near 2.3 $\mu$m and the central region of the CO fundamental band near 4.7 $\mu$m, to compare with similar spectra obtained in 2005-2006.  Our spectrophotometry is consistent with the gradual brightening of the star reported by others. Overall, the near-infrared spectra bear some resemblance to that of a extremely cool supergiant. However, the accurate classification by others of the central star as an M6 giant, based on optical observations, the CO line profiles observed at high resolution, the strong dust continuum observed in the  $M$ band, and the much improved modeling of the the medium resolution 1$-$2.5~$\mu$m spectrum with multiple gas and dust components, demonstrate that in addition to the M6 giant photosphere, much cooler dust and gas outside of the photosphere make major contributions to the infrared spectrum. The velocity profiles of CO fundamental and overtone band lines from low rotational levels of the ground vibrational state have changed since last observed in 2005-2006. However the ejecta resulting from the events of 2002 were continuing to move outward at velocities up to 200 km s$^{-1}$ in 2015.  At the same time the velocities of peak absorption of high $J$ lines of the 2$-$0 band of $^{12}$CO, which are formed in the outer layers of the stellar atmosphere, appear to be slightly redshifted  from the stellar velocity, but by much less than a decade earlier. The profiles of these lines remained asymmetric with extended blue wings. These and the observed brightening suggest that as of 2015 the photosphere of V838 Mon was still disturbed and was continuing to contract, and as of 2022 that the central merged star was continuing to emerge from the dust shell created in 2002.

\section*{Acknowledgements}

The research reported here is based on observations obtained at the Gemini Observatory, a program of NOIRLab,  which is managed by the Association of Universities for Research in Astronomy (AURA) under a cooperative agreement with the National Science Foundation, on behalf of the Gemini Observatory partnership: the National Science Foundation (United States), National Research Council (Canada), Agencia Nacional de Investigaci\'{o}n y Desarrollo (Chile), Ministerio de Ciencia, Tecnolog\'{i}a e Innovaci\'{o}n (Argentina), Minist\'{e}rio da Ci\^{e}ncia, Tecnologia, Inova\c{c}\~{o}es e Comunica\c{c}\~{o}es (Brazil), and Korea Astronomy and Space Science Institute (Republic of Korea). The NASA Infrared Telescope Facility is operated by the University of Hawaii under contract 80HGTR19D0030 with the National Aeronautics and Space Administration. We thank the staffs of the Gemini Observatory and the IRTF for their support. 

\section*{Data Availability}
The raw data in this paper are available from the Gemini Observatory Archive, https://archi ve.gemini.edu/, and from the IRTF archive, http://irtf web.if a.hawaii.edu/research/irtf data archive.php.





\bsp	
\label{lastpage}
\end{document}